\documentclass[review,a4paper,3p,12pt]{elsarticle}

\usepackage[T1]{fontenc}
\usepackage[utf8]{inputenc}
\usepackage{ams math,amssymb}

\usepackage{hyperref}
\usepackage{chngcntr}

\usepackage{graphicx}

\usepackage{natbib}
\usepackage[dvipsnames]{xcolor}
\usepackage{hyperref}
\usepackage{booktabs}
\usepackage{color}
\counterwithout{figure}{section}

\graphicspath{{Images/}}

\begin{document}
	
	\begin{frontmatter}
\title{Cascading failures with group support in interdependent hypergraphs}

\author[inst1]{Lei Chen}
\author[inst,inst4]{Chunxiao Jia}
\author[inst,inst4]{Run-Ran Liu}
\author[inst1,inst4]{Fanyuan Meng}

\affiliation[inst1]{organization={Research Center for Complexity Sciences},
    addressline={Hangzhou Normal University}, 
    city={Hangzhou},
    postcode={311121}, 
    state={Zhejiang},
    country={China}}

\affiliation[inst4]{Corresponding authors: chunxiaojia@163.com; runranliu@163.com; fanyuan.meng@hotmail.com
}
\date{\today}

\begin{abstract}
The functionality of an entity frequently necessitates the support of a group situated in another layer of the system. 
To unravel the profound impact of such group support on a system's resilience against cascading failures, we devise a framework comprising a double-layer interdependent hypergraph system, wherein nodes are capable of receiving support via hyperedges. Our central hypothesis posits that the failure may transcend to another layer when all support groups of each dependent node fail, thereby initiating a potentially iterative cascade across layers.
Through rigorous analytical methods, we derive the critical threshold for the initial node survival probability that marks the second-order phase transition point. A salient discovery is that as the prevalence of dependent nodes escalates, the system dynamics shift from a second-order to a first-order phase transition. Notably, irrespective of the collapse pattern, systems characterized by scale-free hyperdegree distributions within both hypergraph layers consistently demonstrate superior robustness compared to those adhering to Poisson hyperdegree distributions. In summary, our research underscores the paramount significance of group support mechanisms and intricate network topologies in determining the resilience of interconnected systems against the propagation of cascading failures. By exploring the interplay between these factors, we have gained insights into how systems can be designed or optimized to mitigate the risk of widespread disruptions, ensuring their continued functionality and stability in the face of adverse events.
\end{abstract}
\end{frontmatter}
\section{Introduction}
Cascading failures, characterized by the propagation of disruptions across interconnected systems, pose significant threats to various societal structures, including power grids \cite{schafer2018dynamically,song2015dynamic,dey2016impact,kinney2005modeling}, transportation networks \cite{barrett2010cascading,su2014robustness,liu2021cascading,jia2020dynamic}, and communication systems \cite{ren2018stochastic,ghasemi2023robustness,cai2015cascading}. These failures often originate from the malfunction of a single or a small number of entities, ultimately leading to widespread and, at times, catastrophic collapses of entire systems. Such events can disrupt critical services, cause economic losses, and even endanger human lives. Consequently, the study of cascading failures has garnered significant attention from researchers across diverse disciplines, such as physics, mathematics, computer science, and social sciences.

One of the fundamental approaches to studying cascading failures is through the lens of network science \cite{valdez2020cascading,motter2002cascade}. Networks provide a powerful framework for modeling and analyzing the structural properties and relationships between entities within complex systems. By representing entities as nodes and their interactions as edges, networks allow researchers to gain insights into how disruptions propagate through a system and ultimately lead to cascading failures. This network-based representation has proven to be instrumental in unraveling the underlying mechanisms that govern the resilience and vulnerability of various systems.

Furthermore, many real-world systems consist of multiple interdependent layers, where nodes and edges in one layer can depend on those in another \cite{boccaletti2014structure,gao2012networks, havlin2015percolation}. For instance, the functionality of a node in a communication network may rely on the stable operation of a hyperedge in a power grid. This interdependence further complicates the analysis of cascading failures, as disruptions can propagate across layers, triggering failures in multiple systems simultaneously  \cite{buldyrev2010catastrophic,li2012cascading,duan2019universal}. Furthermore, numerous researchers extended the field by introducing link direction \cite{liu2016breakdown}, correlated properties \cite{min2014network,valdez2013triple}, redundant dependencies \cite{radicchi2017redundant}, dependence strength \cite{parshani2010interdependent,dong2012percolation,kong2017percolation}, multiple support \cite{shao2011cascade,gao2011robustness,tong2024cascading} to investigate the robustness of multilayer networks.

However, traditional network models that focus primarily on pairwise interactions between nodes have limitations in accurately representing real-world systems that exhibit higher-order interactions \cite{battiston2020networks,battiston2021physics,benson2016higher}. Higher-order interactions involve simultaneous engagements among groups comprising more than two nodes, which are prevalent in many real-world systems \cite{alvarez2021evolutionary,grilli2017higher}. Notably, hypergraphs provide a more comprehensive framework to model various complex systems for modeling complex systems, capturing the richness and diversity of higher-order interactions due to the fact that hyperedges can connect an arbitrary number of nodes \cite{de2020social,zhang2023higher}.

Despite the increasing recognition of the significance of higher-order interactions and interdependencies in complex systems, the study of cascading failures in interdependent hypergraphs remains relatively limited \cite{chen2024catastrophic,liu2023higher,pan2024robustness,sun2021higher}.  
Importantly, most existing works focus on the dependencies between nodes across different layers, neglecting the possibility that nodes can also depend on different hyperedges (support groups) in another layer. This oversight limits our understanding of how cascading failures propagate in systems with intricate interplay between structural properties and interdependencies.

To address this gap, we propose a novel model to study cascading failures in interdependent hypergraphs. Our model comprises two mutually dependent hypergraphs, each containing a set of nodes and hyperedges. We incorporate the notion of support groups, where nodes in one hypergraph can depend on hyperedges in the other hypergraph for their functionality. By considering both higher-order interactions and interdependencies between nodes and hyperedges across different layers, our model provides a more comprehensive representation of real-world complex systems.

\section{Model Description}
We construct a double-layer interdependent hypergraph system, denoted as $A$ and $B$, comprising $N_A$ and $N_B$ nodes, and $M_A$ and $M_B$ hyperedges, respectively. Each node's hyperdegree, represented by $k$, indicates the number of hyperedges it is part of. The hyperdegree distributions for layers $A$ and $B$ are specified as $P_A(k)$ and $P_B(k)$. Similarly, the cardinality $m$ of a hyperedge, which is the count of nodes it encompasses, adheres to distributions $Q_A(m)$ and $Q_B(m)$.

Our model integrates interdependencies where nodes in one layer rely on support hyperedges from the other. Specifically, a node in layer $A$ is chosen with probability $q_A$ to depend on support hyperedges from layer $B$, and similarly for nodes in layer $B$ with probability $q_B$. The support degree of each dependent node, denoted as $\tilde{k}$, follows distributions $\tilde{P}_A(\tilde{k})$ and $\tilde{P}_B(\tilde{k})$.

The cascading dynamics commence with the removal of nodes with probabilities $1-r_A$ and $1-r_B$ in each respective layer, where $r$ represents the initial node survival probability. Subsequently, a node remains functional only if: (i) it is connected to at least one functional support hyperedge in the other layer; and (ii) it is part of the Giant Connected Component (GCC) within its own layer, ensuring internal connectivity. This cascading failure process, triggered by the initial node removals, iterates between the two layers until no additional nodes fail. Our primary focus is on the size $S$ of the final GCC, which quantifies the system's robustness.

\begin{figure}
    \centering
    \includegraphics[width=\linewidth]{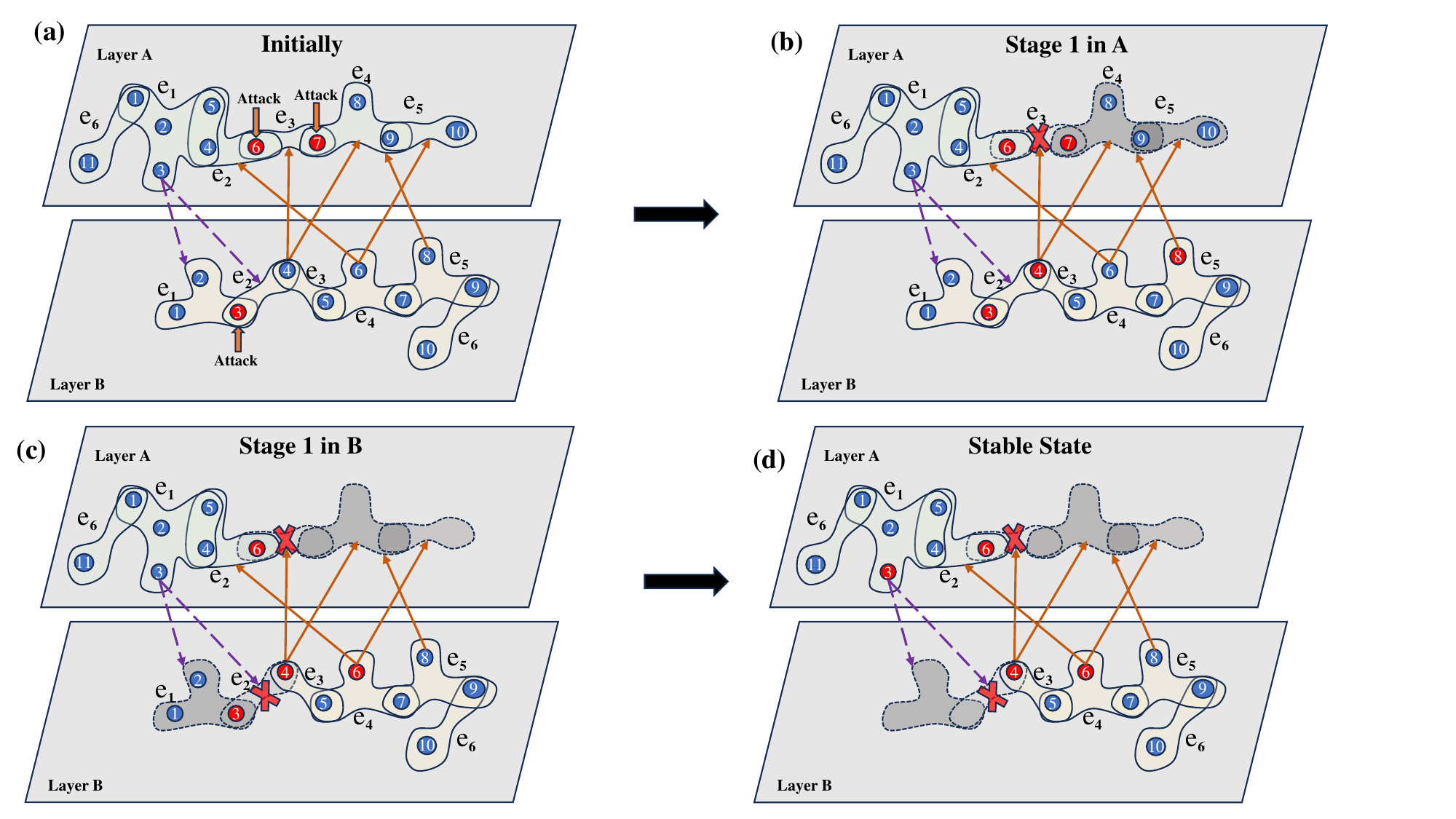}
    \caption{Illustration of a cascading failure process within double-layer hypergraphs $A$ and $B$. Arrow lines signify the interdependencies established between dependent nodes in one layer and their respective support hyperedges in the other layer. Blue nodes indicate functional nodes, contrasting with red nodes, which signify failed nodes. Hyperedges depicted in grey symbolize those that have collapsed due to their disconnection from the Giant Connected Component (GCC), whereas hyperedges marked with red crosses represent failures directly attributed to the collapse of their constituent nodes.}
    \label{fig:fig1}
\end{figure}

The cascading failure process within the model can be illustrated in Fig.~\ref{fig:fig1}. Initially, nodes $6$ and $7$ in layer $A$, along with node $3$ in layer $B$, are removed. At stage $1$ of layer $A$, the hyperedge $e_3$ collapses, which consequently causes the small component consisting of hyperedges $e_4$ and $e_5$ to collapse as it detaches from the GCC. At stage 1 in layer $B$, node $4$ will fail due to the collapse of its support hyperedges $e_3$ and $e_4$ in layer $A$. Furthermore, the collapse of hyperedge $e_2$ in layer $B$, leading to the collapse of the small component that comprises hyperedge $e_1$, as it is disconnected from the GCC of layer $B$. The cascading failure process ceases as no further nodes fail. The final GCC is composed of the component of 
nodes $\{1,2,4,5,11\}$ in layer $A$, and the component of nodes $\{5,7,8,9,10\}$, respectively.

\section{Key Results}
To characterize the structures of both layers, we introduce several generating functions. These functions encapsulate essential information regarding the distribution of hyperdegrees, support degrees, and cardinalities within each layer. The generating functions for the hyperdegree distribution are defined as:
\begin{equation}
\left\{\begin{array}{l}
G_{k0}^{A}(x) = \sum\limits_{k=0}^{\infty} P^A(k) x^{k}, \\
G_{k0}^{B}(x) = \sum\limits_{k=0}^{\infty} P^B(k) x^{k} .
\end{array}\right.
\label{eq:generate_k0}
\end{equation}
The generating functions for the excess hyperdegree distribution are defined as:
\begin{equation}
\left\{\begin{array}{l}
G_{k1}^{A}(x) =  \sum\limits_{k=1}\frac{k P^{A}(k)}{\langle k\rangle} x^{k-1}= G_{k0}^{A\prime}(x) / G_{k0}^{A\prime}(1),\\
G_{k1}^{B}(x) =  \sum\limits_{k=1}\frac{k P^{B}(k)}{\langle k\rangle} x^{k-1} = G_{k0}^{B\prime}(x) / G_{k0}^{B\prime}(1).
\end{array}\right.
\label{eq:generate_k1}
\end{equation}
The generating functions for the support degree distribution are defined as:
\begin{equation}
\left\{\begin{array}{l}
\tilde{G}_{k0}^{A}(x) = \sum\limits_{{k}=0}^{\infty} \tilde{P}^A({k}) x^{{k}}, \\
\tilde{G}_{k0}^{B}(x) = \sum\limits_{{k}=0}^{\infty} \tilde{P}^B({k}) x^{{k}} .
\end{array}\right.
\label{eq:generate_k0_between}
\end{equation}
The generating functions for the cardinality distribution are defined as:
\begin{equation}
\left\{\begin{array}{l}
G_{m0}^{A}(x) = \sum\limits_{m=0}^{\infty} Q^A(m) x^{m}, \\
G_{m0}^{B}(x) = \sum\limits_{m=0}^{\infty} Q^B(m) x^{m} .
\end{array}\right.
\label{eq:generate_m0}
\end{equation}
The generating functions for the excess cardinality distribution are defined as:
\begin{equation}
\left\{\begin{array}{l}
G_{m1}^{A}(x) =  \sum\limits_{m=1}\frac{m Q^{A}(m)}{\langle m\rangle} x^{m-1}=G_{m0}^{A\prime}(x) / G_{m0}^{A\prime}(1), \\
G_{m1}^{B}(x) =  \sum\limits_{m=1}\frac{m Q^{B}(m)}{\langle m\rangle} x^{m-1} = G_{m0}^{B\prime}(x) / G_{m0}^{B\prime}(1).
\end{array}\right.
\label{eq:generate_m1}
\end{equation}

As illustrated in Fig.~\ref{fig:fig1}, the cascading failure process unfolds in discrete iterations, progressing from stage $n=1$ to $n=\infty$ within both layers. 

We denote the probabilities that hyperedges are functional in hypergraphs $A$ and $B$ at stage $n$ as $T^{A}_{n}$ and $T^{B}_{n}$, respectively. Consequently, the probability that a dependent node in hypergraph $A$($B$) at stage $n$ lacks functional support hyperedges in hypergraph $B$($A$) is given by
\begin{equation}
\begin{cases}
    u_n^A=\sum_{{k}=0}^{\infty} \tilde{P}^A({k})\left(1-T_{n-1}^B\right)^{{k}}=\tilde{G}_{k0}^A(1-T_{n-1}^B),\\
    u_n^B=\sum_{{k}=0}^{\infty} \tilde{P}^B({k})\left(1-T_{n-1}^A\right)^{{k}}=\tilde{G}_{k0}^B\left(1-T_{n-1}^A\right).
\end{cases}
\end{equation}

The fraction of nodes in hypergraph $A$($B$) which remain functional at stage $n$ after applying condition (i) is derived as
\begin{equation}
\begin{cases}
    p_n^A=r^A\left(1-q^{A}u_n^A  \right),\\
    p_n^B=r^B\left(1-q^{B}u_n^B \right).
\end{cases}
\end{equation}

Subsequently, the probabilities $T_n^A$ and $T_n^B$ are expressed as
\begin{equation}
\left\{\begin{array}{l}
T_{n}^{A}=\sum\limits_{m=0} Q^{A}(m)\sum\limits_{j=0}^{m}\binom{m}{j}
\left\{ 1-[G_{k1}^{A} \left( 1- f_{n}^{A}\right)]^{m-j} \right\}(p_{n}^{A})^{m-j} \left(1-p_{n}^{A}\right)^{j},\\
T_{n}^{B}=\sum\limits_{m=0} Q^{B}(m)\sum\limits_{j=0}^{m}\binom{m}{j}
\left\{ 1-[G_{k1}^{B} \left( 1- f_{n}^{B}\right)]^{m-j} \right\}(p_{n}^{B})^{m-j} \left(1-p_{n}^{B}\right)^{j},\\  
\end{array}\right.
\label{eq:T_n_A_B}
\end{equation}
where $f_{n}^{A}$ or $f_{n}^{B}$ represents the probability that 
a randomly selected hyperedge, reached through a random node, can connect to the
GCC of each layer (condition (ii)). The term $1-[G_{1}^{A} \left( 1- f_{n}^{A}\right)]^{m-j}$ or $1-[G_{1}^{B} \left( 1- f_{n}^{B}\right)]^{m-j}$ denotes the probability that within a random hyperedge of cardinality $m$, at least one node out of $m-j$ functional nodes along the remaining hyperedges can connect to the GCC. 

Analogously, we can derive $f_{n}^{A}$ and $f_{n}^{B}$ as 
\begin{equation}
\left\{\begin{array}{l}
f_{n}^{A}=\sum\limits_{m=1}\frac{m Q^{A}(m)}{\langle m\rangle}\sum\limits_{j=0}^{m-1}\binom{m-1}{j}
\left\{ 1-[G_{k1}^{A} \left( 1- f_{n}^{A}\right)]^{m-1-j} \right\}(p_{n}^{A})^{m-1-j} \left(1-p_{n}^{A}\right)^{j} ,\\
f_{n}^{B}=\sum\limits_{m=1}\frac{m Q^{B}(m)}{\langle m\rangle}\sum\limits_{j=0}^{m-1}\binom{m-1}{j}
\left\{ 1-[G_{k1}^{B} \left( 1- f_{n}^{B}\right)]^{m-1-j} \right\}(p_{n}^{B})^{m-1-j} \left(1-p_{n}^{B}\right)^{j}.
\end{array}\right.
\label{eq:f_n_A_B}
\end{equation}

Therefore, the fractions $S_n^A$ and $S_n^B$ of nodes in the GCC relative to the original sizes can be obtained as
\begin{equation}
\left\{\begin{array}{l}
        S_{n}^{A}=p_{n}^{A}[1- G_{k0}^{A}(1-f_{n}^{A})],\\
        S_{n}^{B}=p_{n}^{B}[1- G_{k0}^{B}(1-f_{n}^{B})].
\end{array}\right.
\end{equation}

Upon termination of the cascading failure process, $p_n^A$, $p_n^B$, $T_{n}^{A}$, $T_{n}^{B}$, $f_{n}^{A}$, $f_{n}^{B}$, $S_{n}^{A}$, and $S_{n}^{B}$ all converge to their steady values $p_{\infty}^A$, $p_{\infty}^B$, $T_{\infty}^{A}$, $T_{\infty}^{B}$, $f_{\infty}^{A}$, $f_{\infty}^{B}$, $S_{\infty}^{A}$, and $S_{\infty}^{B}$, respectively. 

Based on the generating functions of Eqs.~\eqref{eq:generate_m0} and \eqref{eq:generate_m1}, the steady values of $T_\infty^A$ from Eq.~\eqref{eq:T_n_A_B} and $f_\infty^A$ from Eq.~\eqref{eq:f_n_A_B} can be simplified into 
 \begin{equation}
\left\{\begin{array}{l}
T_{\infty}^{A}=
1
-G_{m0}^{A}\left(1-p_{\infty}^{A}  + p_{\infty}^{A}G_{k1}^{A} \left( 1- f_{\infty}^{A}\right)\right),
\\
f_{\infty}^{A}=
 1
-G_{m1}^{A}\left(1-p_{\infty}^{A}  + p_{\infty}^{A}G_{k1}^{A} \left( 1- f_{\infty}^{A}\right)\right) .
\end{array}\right.
\label{eq:T_f_A_infinite}
\end{equation}

Similarly, we can obtain 
\begin{equation}
\left\{\begin{array}{l}
T_{\infty}^{B}=
1
-G_{m0}^{B}\left(1-p_{\infty}^{B}  + p_{\infty}^{B}G_{k1}^{B} \left( 1- f_{\infty}^{B}\right)\right),
\\
  f_{\infty}^{B}=
1
-G_{m1}^{B}\left(1-p_{\infty}^{B}  + p_{\infty}^{B}G_{k1}^{B} \left( 1- f_{\infty}^{B}\right)\right) .
\end{array}\right.
\label{eq:T_f_B_infinite}
\end{equation}

Furthermore, the final fractions $S_\infty^A$ and $S_\infty^B$ of nodes in the GCC can be expressed as
\begin{equation}
\left\{\begin{array}{l}
        S_{\infty}^{A}=p_{\infty}^{A}[1- G_{k0}^{A}(1-f_{\infty}^{A})],\\
        S_{\infty}^{B}=p_{\infty}^{B}[1- G_{k0}^{B}(1-f_{\infty}^{B})].
\end{array}\right.
\end{equation}

\section{Benchmark: symmetric hypergraphs}
Given the complexity of Eqs.~\eqref{eq:T_f_A_infinite} and \eqref{eq:T_f_B_infinite}, which can be solved only numerically for most cases, we first assume that both layers follow a Poisson cardinality distribution, i.e., $\frac{e^{-{\langle m \rangle}} {\langle m \rangle}^m}{m!}$. Consequently, we have 
\begin{equation}
    \begin{cases}
        G_{m0}^A(x)= G_{m1}^A(x),\\
        G_{m0}^B(x)= G_{m1}^B(x).
    \end{cases}
\end{equation}

From Eqs.~\eqref{eq:T_f_A_infinite} and \eqref{eq:T_f_B_infinite}, we derive 
\begin{equation}
    \begin{cases}
        f_{\infty}^{A} =T_{\infty}^{A},\\
        f_{\infty}^{B} =T_{\infty}^{B}.
    \end{cases}
\end{equation}

This simplifies Eqs.~\eqref{eq:T_f_A_infinite} and \eqref{eq:T_f_B_infinite} to
 \begin{equation}
\left\{\begin{array}{l}
  f_{\infty}^{A}=
1
-G_{m0}^{A}\left(1-p_{\infty}^{A}  + p_{\infty}^{A}G_{k1}^{A} \left( 1- f_{\infty}^{A}\right)\right), 
\\
f_{\infty}^{B}=
1
-G_{m0}^{B}\left(1-p_{\infty}^{B}  + p_{\infty}^{B}G_{k1}^{B} \left( 1- f_{\infty}^{B}\right)\right).
\end{array}\right.
\label{eq:f_A_B_infinite}
\end{equation}

Moreover, if both hypergraphs exhibit identical cardinality, hyperdegree, and support degree distributions, i.e., symmetric hypergraphs, we obtain
\begin{equation}
    \begin{cases}
        G_{m0}^A(x) = G_{m0}^B(x)=G_{m0}(x),\\
        G_{k0}^A(x) = G_{k0}^B(x)=G_{k0}(x),\\
        \tilde{G}_{k0}^A(x) = \tilde{G}_{k0}^B(x) =\tilde{G}_{k0}(x).
    \end{cases}
\end{equation}

Assuming identical parameter settings, i.e., $q_A = q_B = q$, $r^A = r^B=r$, we find
\begin{equation}
    f_{\infty}^{A}=f_{\infty}^{B}= f. 
\end{equation}

Thus, Eq.~\eqref{eq:f_A_B_infinite} simplifies to
\begin{equation}
f=1 - G_{m0}\left(1-r[1-q\widetilde{G}_{k0}(1-f)][1-  G_{k1} \left( 1- f\right)]\right).
\label{eq:f}
\end{equation}

Additionally, the size of GCC, $S_{\infty}^{A}=S_{\infty}^{B} = S$, is given by
\begin{equation}
    S=r[1-q\widetilde{G}_{k0}(1-f)][1- G_{k0}(1-f)].
    \label{eq:S}
\end{equation}
\subsection{The critical point}
A critical fraction of dependent nodes, $q_c$, exists, below which a second-order phase transition occurs, and above which a first-order phase transition occurs.

To solve the critical point $r_c^{II}$ for the second-order phase transition when $q< q_c$, we define
\begin{equation}
	h(r,f) = 1 - G_{m0}\left(y(f)\right)  - f,
	\label{eq:hf}
\end{equation}
with
\begin{equation}
    y(f) = 1-r[1-q\widetilde{G}_{k0}(1-f)][1-  G_{k1} \left( 1- f\right)].
\end{equation}

Since $r_c^{II}$ satisfies $\partial_f h(r_c^{II},0)=0$, we derive
\begin{equation}
    G_{m0}^\prime(y(0)) y ^ \prime(0) - 1 = 0,
\end{equation}
with 
\begin{equation}
    \begin{cases}
        y(0) = 1,\\
        y'(0) = {G}_{k1}^{\prime}(1)(1-q).
    \end{cases}
\end{equation}

Given that both layers follow Poisson cardinality distributions, we obtain
\begin{equation}
	r_c^{II}  = \frac{1}{\langle m \rangle(1-q) \frac{\langle k^2 \rangle - \langle k \rangle }{\langle k \rangle}}.
	\label{eq:r2_all}
\end{equation}

Furthermore, if both layers follow a Poisson hyperdegree distribution, we derive
\begin{equation}
	r_c^{II}  = \frac{1}{\langle m \rangle(1-q) \langle k \rangle}.
	\label{eq:r2_er}
\end{equation}

\subsection{The triple point}

\begin{figure}[h]
	\centering
	\includegraphics[width=\linewidth]{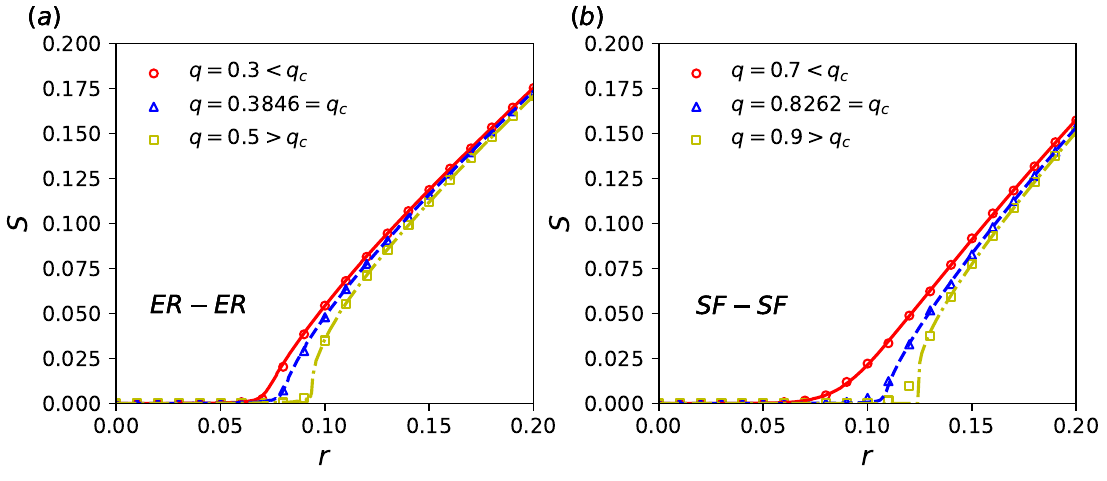}
	\caption{The fraction $S$ of GCC as a function of $r$ for varying $q$. The support degree and cardinality are modeled by a Poisson distribution with $\langle \tilde{k} \rangle=4$ and $\langle m \rangle=5$, respectively. (a) illustrates the scenario where the hyperdegree adheres to a Poisson distribution (ER) with $\langle {k} \rangle=4$; (b) depicts the case where the hyperdegree follows a scale-free power-law distribution $p(k) \sim  k^{-\lambda}$ (SF), characterized by $\lambda = 2.55$, with a minimum hyperdegree $k_{min}=2$ and a maximum hyperdegree $k_{max}=100$. The lines denote theoretical predictions, while the points represent simulation results obtained with $N=10^4$ nodes and $M=8 \times 10^3$ hyperedges.}	
	\label{fig:fig2}
\end{figure}

The fraction $q$ of dependent nodes has a great impact on the collapse pattern of the system. As illustrated in Fig.~\ref{fig:fig2}, a pivotal transition in the system's collapse pattern is triggered as the value of $q$ exceeds the triple point $q_c$.
This transition, from a smooth, second-order phase transition to an abrupt, first-order phase transition, signifies a profound shift in the system's resilience.

Since at the critical point $q_c$, the condition for second-order and first-order phase transition are met, we must have the equation $\partial^2_f h(r_c^{II},0)=0$ which yields
\begin{equation}
    G_{m0}^{''}(y(0)) (y'(0))^2 + G_{m0}^{'}(y(0)) y''(0) = 0,
    \label{eq:second_order_de}
\end{equation}
with
\begin{equation}
\begin{cases}
    y(0) = 1, \\
    y'(0) = -r_c^{II} (1-q_c) G_{k1}^\prime (1) ,\\
    y''(0) = -r_c^{II} \left(2 q_c \tilde{G_{k0}^\prime}(1) G_{k1}^{'} (1) - (1-q_c) G_{k1}^{''} (1) \right).
\end{cases}
\end{equation}

Since both layers follow a cardinality Poisson distribution, we further get
\begin{equation}
	q_c= \frac{1}{1+\frac{2{G}_{k1}^{\prime}(1){\tilde{G}}_{0}^{\prime}(1)}{{G}_{k1}^{\prime}(1)+{G}_{k1}^{\prime\prime}(1)}}.
	\label{eq:qc_all}
\end{equation}

Additionally, if hyperdegree and support degree follow a Poisson distribution for both layers, we can obtain
\begin{equation}
	q_c= \frac{1}{1+\frac{2\langle \tilde{k} \rangle}{\langle k \rangle+1}}.
	\label{eq:q_c_Poisson}
\end{equation}

\begin{figure}
	\centering
	\includegraphics[width=\linewidth]{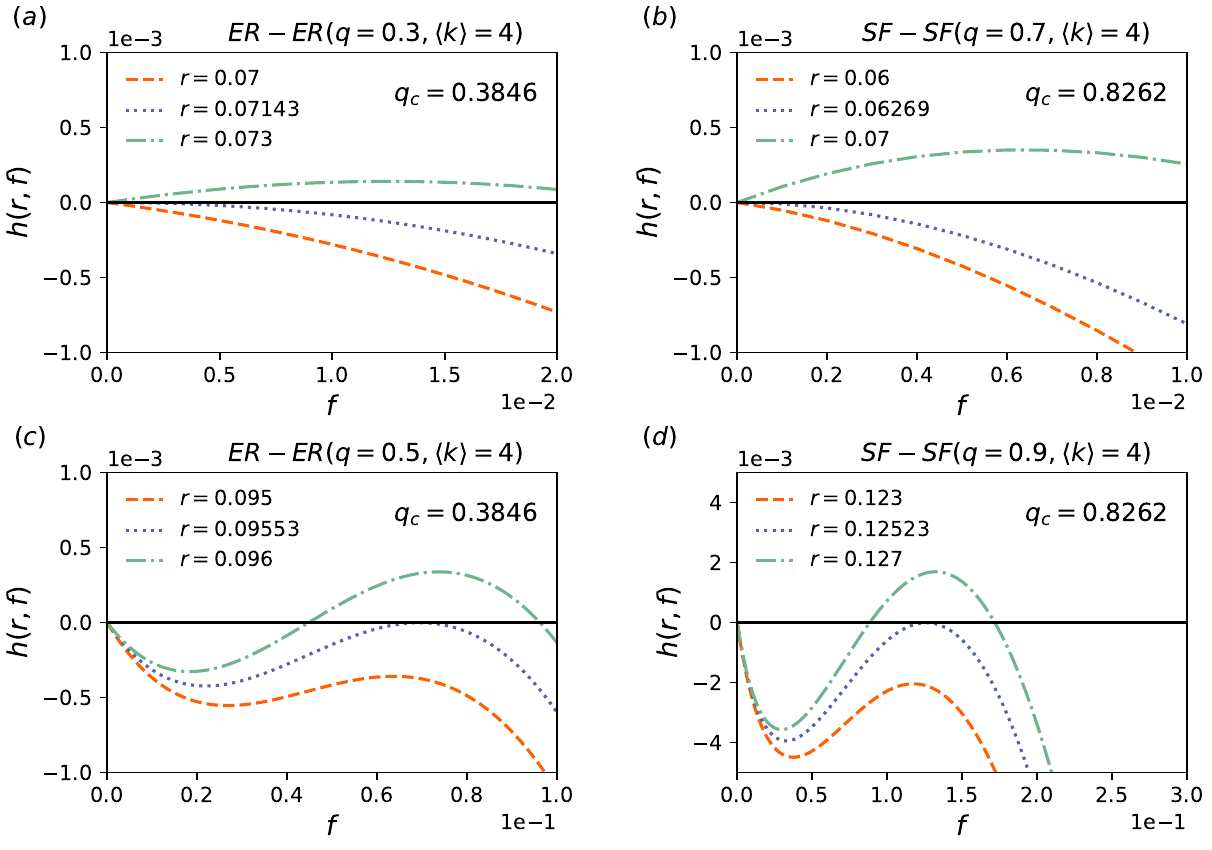}
	\caption{Graphical solution of Eq.~\eqref{eq:hf}. The support degree and cardinality are modeled by a Poisson distribution with $\langle \tilde{k} \rangle=4$ and $\langle m \rangle=5$, respectively. (a,c) illustrates the scenario where the hyperdegree adheres to a Poisson distribution (ER) with $\langle {k} \rangle=4$ for $q=0.3 < q_c = 0.3846$ and $q=0.5 > q_c = 0.3846$, respectively; (b,d) depicts the case where the hyperdegree follows a scale-free power-law distribution $p(k) \sim  k^{-\lambda}$ (SF), characterized by $\lambda = 2.55$, with a minimum hyperdegree $k_{min}=2$ and a maximum hyperdegree $k_{max}=100$ for $q=0.7 < q_c = 0.8262$ and $q=0.9 > q_c = 0.8262$, respectively.}
	\label{fig:fig3}
\end{figure}

In Fig.~\ref{fig:fig2}(a), the curve labeled by $q = 0.3846 = q_c$ (acquired by Eq.~\eqref{eq:q_c_Poisson}), delineates the boundary between second-order and first-order phase transition. For $q=0.3 < q_c$, the system undergoes a second-order phase transition at the critical point $r_c^{II} \approx 0.07143$ (verfied by Fig.~\ref{fig:fig3}(a)). Conversely, when $q=0.5>q_c$, the collapse pattern of the system is an abrupt first-order phase transition at the critical point $r_c^{I} \approx 0.09553$ (verified by Fig.~\ref{fig:fig3}(b)). These observations remain consistent when the hypergraphs adhere to a scale-free hyperdegree distribution, as seen in Fig.~\ref{fig:fig2}(b), thereby underscoring the robustness of our results across varying hyperdegree distributions.


\begin{figure}
    \centering    \includegraphics[scale=0.6]{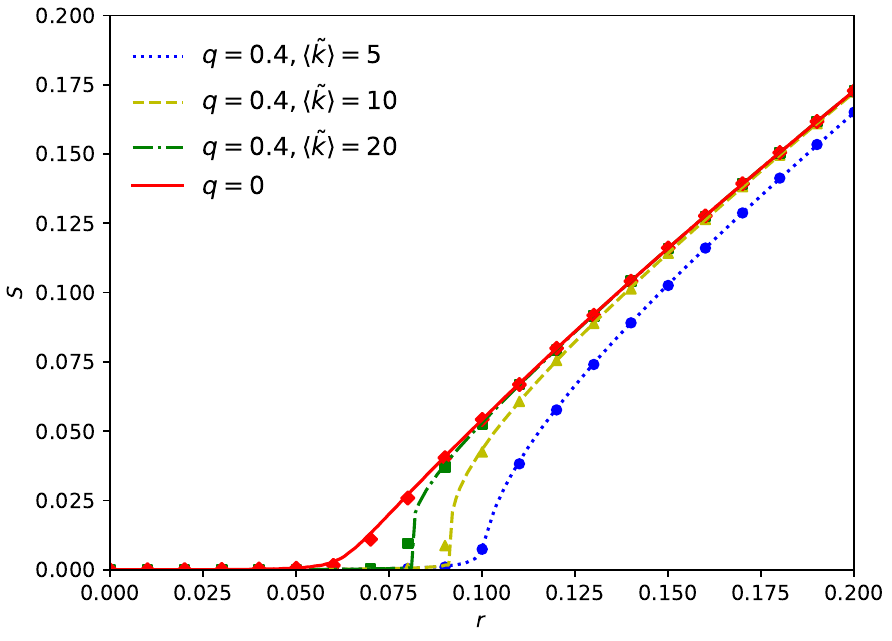}
    \caption{The fraction $S$ of GCC as a function of $r$ for varying $\langle \tilde{k} \rangle$. The hyperdegree and cardinality are modeled by a Poisson distribution with $\langle k \rangle=4$ and $\langle m \rangle=4$, respectively. The lines denote theoretical predictions, while the points represent simulation results obtained with $N=10^4$ nodes and $M=10^4$ hyperedges.}
    \label{fig:transition_q_c}
\end{figure}

Moreover, Eq.~\eqref{eq:q_c_Poisson} reveals a
profound relationship: the triple point $q_c$ is inversely proportional to the average support degree $\langle \tilde{k} \rangle$ and directly proportional to the average hyperdegree $\langle {k}\rangle$. This relationship highlights a crucial principle: an augmentation in the density of support hyperedges within the system triggers a shift towards a lower transition threshold separating second-order from first-order phase transitions. Specifically, in Fig.~\ref{fig:transition_q_c}, when $\langle \tilde{k} \rangle=5$, $q=0.4<q_c$ results in the system undergoing a second-order phase transition. However, as $\langle \tilde{k} \rangle$ increases to 10, $q=0.4>q_c$ occurs, causing the system to exhibit first-order phase transition. It is important to note that, a higher value of $\langle \tilde{k} \rangle$ makes the system more robust, the potential for abrupt collapse necessitates vigilance.

\begin{figure}
    \centering    \includegraphics[scale=0.6]{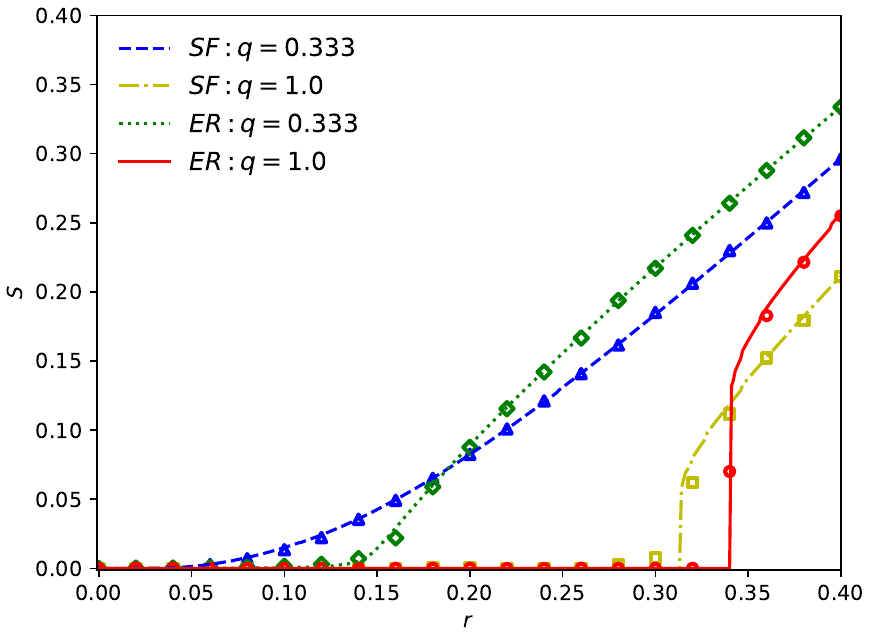}
    \caption{The fraction $S$ of GCC as a function of $r$ for varying $q$. The support degree and cardinality are modeled by a Poisson distribution with $\langle \tilde{k} \rangle=4$ and $\langle m \rangle=5$, respectively. 
    This illustration compares the robustness of a system where both layers follow a Poisson hyperdegree distribution (ER) with $\langle {k} \rangle=5$ against a system where both layers follow a scale-free power-law distribution $p(k) \sim  k^{-\lambda}$ (SF) with $\lambda = 2.31$, $k_{min}=2$ and $k_{max}=136$. The lines denote theoretical predictions, while the points represent simulation results obtained with $N=10^4$ nodes and $M=10^4$ hyperedges.}
    \label{fig:ER-SF}
\end{figure}

Lastly, it is important to highlight that, irrespective of the collapse patterns, a system consisting of two layers with scale-free hyperdegree distributions consistently exhibits greater robustness than systems with Poisson hyperdegree distributions (see Fig.~\ref{fig:ER-SF}).

\newpage
\section{Conclusion}
In conclusion, the study of cascading failures within interdependent systems is crucial due to their widespread presence across diverse infrastructures. These failures, capable of rapid propagation through interconnected layers, present significant challenges to the functionality and stability of entire systems. Our research fills a critical void in the literature by exploring cascading failures in interdependent hypergraphs, which more accurately capture the higher-order interactions found in real-world systems. By incorporating both higher-order interactions and interdependencies between nodes and hyperedges across layers, our model offers a comprehensive framework to dissect the mechanisms that govern the resilience of complex systems against cascading failures.

Our findings yield several profound insights into the resilience of interdependent hypergraph systems. Initially, we analytically determine a critical threshold for the initial node survival probability, which delineates the second-order phase transition in system dynamics. Moreover, we analytically derive the critical fraction of dependent nodes, which signifies the transition between second-order and first-order phase transitions. This triple point is intimately linked to the support degree and the hyperdegree. Additionally, we observed that systems with scale-free hyperdegree distributions consistently demonstrate enhanced robustness compared to those with Poisson hyperdegree distributions, regardless of the collapse pattern. This highlights the pivotal role of network topology in shaping the resilience of interconnected systems.

Although our model provides valuable insights into cascading failures in interdependent hypergraphs, several avenues for future research remain open. For instance, we could introduce dependence within each hyperedge, where the failure of a node could result in the collapse of the entire hyperedge it is part of. Furthermore, our model currently considers only two interdependent layers; expanding the framework to multi-layer systems could unveil more intricate patterns of failure propagation. Additionally, while we focused on the impact of hyperdegree and support degree distributions, other structural attributes, such as clustering coefficients and community structures, may also play a significant role in the resilience of interdependent hypergraphs.

In summary, our research highlights the profound importance of group support mechanisms and complex network topologies in determining the resilience of interconnected systems against cascading failures. By developing a novel framework that encompasses a double-layer interdependent hypergraph system, we have gained insights into how failures propagate across layers and culminate in system-wide collapses. Our findings, including the identification of critical thresholds and the superiority of scale-free hyperdegree distributions, offer valuable guidance for designing and optimizing complex systems to mitigate the risk of widespread disruptions. While our model marks a significant advancement, numerous opportunities exist for future research to further refine and expand our understanding of cascading failures in interdependent hypergraphs.

\section{Data Availability}
 Data is available on request from the authors. 
 \section{Conflict of Interest Statement}
 The authors declare no conflict of interest.
 \section{Acknowledgements}
     This work is supported by the National Natural Science Foundation of China (Grant No.61773148 and No.52374013), and the Entrepreneurship and Innovation Project of High Degree Returned Overseas Scholar in Hangzhou.
\newpage
\bibliographystyle{unsrt}
\bibliography{main.bib}
\end{document}